\documentclass[english]{llncs}
\usepackage[T1]{fontenc}
\usepackage[latin9]{inputenc}
\usepackage{array}
\usepackage{verbatim}
\usepackage{latexsym}
\usepackage{units}
\usepackage{amsmath}
\usepackage{amssymb}
\usepackage{stmaryrd}
\usepackage{setspace}
\makeatletter

\providecommand{\tabularnewline}{\\}

\usepackage{xcolor}
\definecolor{darkred}{rgb}{0.8,0,0} 
\usepackage{hyperref}
\hypersetup{final, colorlinks=true, linkcolor=darkred, urlcolor=darkred, citecolor=black, linkbordercolor=darkred, pdfborderstyle={/S/U/W 1}}
\usepackage{setspace}
\usepackage[absolute,overlay]{textpos} 
\usepackage{listings}
\usepackage{cite}
\usepackage{tikz}
\usetikzlibrary{automata,arrows,shapes,decorations,topaths,trees,backgrounds,shadows}

\DeclareRobustCommand{\VAN}[2]{#2}

\usepackage{multicol}

\AtBeginDocument{
  
}

\makeatother

\usepackage{babel}
\begin{document}
\global\long\def\seq{\boldsymbol{x_{1}},\boldsymbol{x_{2}},...}

\global\long\def\foo{\operatorname{foo}}

\setstretch{1}
\title{Turing Completeness of \\Finite, Epistemic Programs}
\author{Dominik Klein\thanks{Department of Philosophy, Bayreuth University, and Department of Political Science, University of Bamberg.  \email{dominik.klein@uni-bayreuth.de}}
and Rasmus K. Rendsvig\thanks{Theoretical Philosophy, Lund University,  and Center for Information and Bubble Studies, University of Copenhagen. \email{rendsvig@gmail.com}}
\institute{}}
\maketitle

\noindent In this note, we present the proof of Lemma 1.1 of \cite{KleinRendsvig-LORI2017},
namely that the class of \textbf{epistemic programs} \cite{BaltagMoss2004}
is \textbf{Turing complete}. Following preliminary definitions in
Section \ref{sec:Definitions}, Section \ref{sec:Theorem-and-Proof}
states and proves the theorem.

\section{Definitions\label{sec:Definitions}}

Let there be given a countable set $\Phi$ of \textbf{atoms} and a
finite set $I$ of \textbf{agents}. Where $p\in\Phi$ and $i\in I$,
define the \textbf{language} $\mathcal{L}$ by
\begin{center}
$\varphi:=\top\;|\;p\;|\;\neg\varphi\;|\;\varphi\wedge\varphi\;|\;\square_{i}\varphi$.
\par\end{center}

We use relational semantics to evaluate formulas. A \textbf{Kripke
model} for $\mathcal{L}$ is a tuple $M=(\left\llbracket M\right\rrbracket ,R,\left\llbracket \cdot\right\rrbracket )$
where $\left\llbracket M\right\rrbracket $ is a countable, non-empty
set of \textbf{states}, $R:I\longrightarrow\mathcal{P}(\left\llbracket M\right\rrbracket \times\left\llbracket M\right\rrbracket )$
assigns to each $i\in I$ an \textbf{accessibility relation} $R_{i}$,
and $\left\llbracket \cdot\right\rrbracket :\Phi\longrightarrow\mathcal{P}(\left\llbracket M\right\rrbracket )$
is a \textbf{valuation}, assigning to each atom a set of states. With
$s\in\left\llbracket M\right\rrbracket $, call $Ms=(\left\llbracket M\right\rrbracket ,R,\left\llbracket \cdot\right\rrbracket ,s)$
a \textbf{pointed Kripke model}. The used semantics are standard (see
e.g. \cite{BlueModalLogic2001,ModelTheoryModalLogic}), including
the modal clause:
\[
Ms\models\square_{i}\varphi\mbox{ iff for all }t:sR_{i}t\mbox{ implies }Mt\models\varphi.
\]

Pointed Kripke models may be updated using \textbf{action models }and
\textbf{product update} \cite{BaltagBMS_1998,BaltagMoss2004,Benthem2006_com-change,Ditmarsch2008,Ditmarsch_Kooi_ontic}.
We here invoke a set of mild, but non-standard, requirements to fit
the framework of \cite{KleinRendsvig-LORI2017}.\medskip{}

A \textbf{multi-pointed action model} is a tuple $\Sigma{\scriptstyle \Gamma}=(\llbracket\Sigma\rrbracket,\mathsf{R},pre,\Gamma)$
where $\left\llbracket \Sigma\right\rrbracket $ is a countable, non-empty
set of \textbf{actions}. The map $\mathsf{R}:I\rightarrow\mathcal{P}(\left\llbracket \Sigma\right\rrbracket \times\left\llbracket \Sigma\right\rrbracket )$
assigns an \textbf{accessibility relation} $\mathsf{R}(i)$ on $\Sigma$
to each agent $i\in I$. The map $pre:\left\llbracket \Sigma\right\rrbracket \rightarrow\mathcal{L}$
assigns to each action a \textbf{precondition}. Finally, $\emptyset\not=\Gamma\subseteq\left\llbracket \Sigma\right\rrbracket $
is the set of designated actions. 

Where $X$ is a set of pointed Kripke models, call $\Sigma{\scriptstyle \Gamma}$
\textbf{deterministic} if $\models pre(\sigma)\wedge pre(\sigma')\rightarrow\bot$
for each $\sigma\neq\sigma'\in\Gamma$. 

Let $\Sigma{\scriptstyle \Gamma}$ be deterministic over $X$ and
let $Ms\in X$. Then the \textbf{product update} of $Ms$ with $\Sigma{\scriptstyle \Gamma}$,
denoted $Ms\otimes\Sigma{\scriptstyle \Gamma}$, is the pointed Kripke
model $(\left\llbracket M\Sigma\right\rrbracket ,R',\llbracket\cdot\rrbracket',s')$
with
\begin{eqnarray*}
\left\llbracket M\Sigma\right\rrbracket  & = & \left\{ (s,\sigma)\in\left\llbracket M\right\rrbracket \times\left\llbracket \Sigma\right\rrbracket :(M,s)\models pre(\sigma)\right\} \\
R' & = & \left\{ ((s,\sigma),(t,\tau)):(s,t)\in R_{i}\mbox{ and }(\sigma,\tau)\in\mathsf{R}_{i}\right\} ,\text{ for all }i\in N\\
\left\llbracket p\right\rrbracket ' & = & \left\{ (s,\sigma)\!:\!s\in\left\llbracket p\right\rrbracket \right\} ,\text{ for all }p\in\Phi\\
s' & = & (s,\sigma):\sigma\in\Gamma\mbox{ and }Ms\models pre(\sigma)
\end{eqnarray*}

\noindent As \textsc{$\Sigma{\scriptstyle \Gamma}$ }is assumed deterministic
over $X$ at most one suitable $s'$ exists. If $Ms\models\neg pre(\sigma)$
for all $\sigma\in\Gamma$, $Ms\otimes\Sigma{\scriptstyle \Gamma}$
is undefined.

\section{Theorem and Proof\label{sec:Theorem-and-Proof}}

\noindent Call a finite, deterministic multi-pointed action an \textbf{epistemic
program}.\footnote{The term stems from the seminal \cite{BaltagMoss2004}.} We then show:
\begin{theorem}
\label{thm}The set of epistemic programs is Turing complete. 
\end{theorem}

\begin{remark}
\noindent The proof uses a strict sub-class of the mentioned action
models, all with only equivalence relations as suited for multi-agent
$S5$ logics, and requires only the use of finite, $S5$ pointed Kripke
models. \qed
\end{remark}

\subsubsection{Preliminaries.}

Define a \textbf{Turing machine }as a 7-tuple
\[
\mathsf{M}=(Q,q_{0},q_{h},\Gamma,b,\Sigma,\delta)
\]
where $Q$ is a finite set of \textbf{states} with $q_{0}\in Q$ the
\textbf{start state} and $q_{h}\in Q$ the \textbf{halt state}, $\Gamma$
a finite set of \textbf{tape symbols} with $b\in\Gamma$ the \textbf{blank
symbol} and $\Sigma=\Gamma\backslash\{b\}$ the set of \textbf{input
symbols}, and $\delta$ a partial function
\[
\delta:Q\times\Gamma\rightarrow Q\times\Gamma\times\{l,h,r\}
\]
with $\delta(q_{h},\gamma)$ undefined for all $\gamma\in\Gamma$,
called the \textbf{transition function}. If $\delta(q,\gamma)$ is
undefined, the machine will halt. 

A Turing machine acts on a bi-infinite \textbf{tape} with cells indexed
by $\mathbb{Z}$ and labeled with $\Gamma$ such that only $b$ occurs
on the tape infinitely often. With the machine in state $q\in Q$
and reading label $\gamma\in\Gamma$, the transition function determines
a possibly new state of the machine $q'\in Q$, a symbol $s'$ to
replace $s$ at the current position on the tape, and a movement of
the metaphorical ``read/write head'': Either one cell to the left
($l$), none (stay here, $h$), or one cell to the right ($r$).

A\textbf{ configuration} of a machine is fully given by \emph{i)}
the current labeling of the tape, \emph{ii)} the position of the $\nicefrac{r}{w}$-head
on the tape, and \emph{iii)} the state of the machine.  The space
of possible configurations of a machine $\mathsf{M}$ is thus $\mathfrak{C}=\mathfrak{T}\times\mathbb{Z}\times Q$,
where $\mathfrak{T}$ is the set of bi-infinite strings $\mathfrak{t}=(\dots,\gamma_{-2},\gamma_{-1},\gamma_{0},\gamma_{1},\gamma_{2},\dots)$
over $\Gamma$ such that only $b$ occurs infinitely often in $\mathfrak{t}$.
The transition function $\delta$ of $\mathsf{M}$ may thus be recast
as a partial function $\delta:\mathfrak{C}$$\rightarrow\mathfrak{C}$.

We want to recast $\delta$ in a slightly different manner. Each tape
$\mathfrak{t}$ has infinite head and tail consisting solely of $b$s.
Ignoring all but a finite segment of these yields a finite non-unique
representation of the tape. Formally, for a string $\mathfrak{t}=(\dots,\gamma_{-2},\gamma_{-1},\gamma_{0},\gamma_{1},\gamma_{2},\dots)$
and $k<k'$ let $\mathfrak{t}_{\upharpoonright[k,k']}$ be the substring
$(\gamma_{k},\ldots,\gamma_{k'})$. The set of all such finite representations
of $\mathfrak{T}$ is then given by $T=\{t=(\gamma_{k},\dots,\gamma_{k'})\colon\exists\mathfrak{t}\in\mathfrak{T}\text{ s.t. }t\mathfrak{=t}_{\upharpoonright[k,k']}\text{ and }\forall j<k,\forall j'>k',\mathfrak{t}_{j}=\mathfrak{t}_{j'}=b\}.$
Each $t\in T$ corresponds to a unique $\mathfrak{t}\in\mathfrak{T}$.
Conversely,  each configuration $\mathfrak{c}=(\mathfrak{t},i,q)\in\mathfrak{C}$
may be represented by the equivalence class $\{\mathfrak{(t}_{\upharpoonright[k,k']},i,q)\colon k<k'\}$
of its finite approximations. In each such equivalence class, there
exists representatives for which the position $i$ of the read-write
head is ``on the tape'', i.e., satisfies that $\gamma_{i}\in t$.
We impose this as a requirement and define a restricted equivalence
class for each $\mathfrak{c}=(\mathfrak{t},i,q)\in\mathfrak{C}$ by
$[\mathfrak{c}]=\{\mathfrak{(t}_{\upharpoonright[k,k']},i,q)\colon k\leq i\leq k'\}$.
With $\mathsf{C}=\{[\mathfrak{c}]\colon\mathfrak{c}\in\mathfrak{C}\}$,
i.e., the set of equivalence classes of finite representations of
configurations for which the read-write head is on the finite tape,
the transition function may finally be recast as a partial function
$\delta:\mathsf{C}\rightarrow\mathsf{C}$.

\begin{remark}
The class of Turing machines with $\Gamma=\{0,1\},b=0,$ is Turing
complete.  Henceforth, we restrict attention to this sub-class.\qed
\end{remark}

\subsection*{Proof}

To prove Theorem \ref{thm}, it must be shown that any Turing machine
can be simulated by an epistemic program. We show that as follows:
First, we define an invertible operator $\mathtt{K}$ that for any
finite representation of a configuration $\mathsf{c}\in[\mathfrak{c}]\in\mathsf{C}$
produces a pointed Kripke model $\mathtt{K}(\mathsf{c})$. Second,
we define an epistemic program $\Sigma{\scriptstyle \Gamma}$ which
satisfies that 
\begin{equation}
\mathtt{K}^{-1}(\mathtt{K}(\mathsf{c})\otimes\Sigma{\scriptstyle \Gamma})\in\delta([\mathfrak{c}]),\label{this}
\end{equation}
for any $[\mathfrak{c}]\in\mathsf{C}$. Hence $\Sigma{\scriptstyle \Gamma}$ may be used to
calculate the trajectory of $\delta$.

\subsubsection{Machine, Language and Logic.}

\noindent Fix a Turing machine $\mathsf{M}$ with states $Q$, and
fix from this a set of relation indices $Q'=Q\cup\{a,b,1\}$. Let
the modal language $\mathcal{L}$ be based on the single atom $p$
and operators $\square_{i},i\in Q'$.

\subsubsection{Configuration Space.}

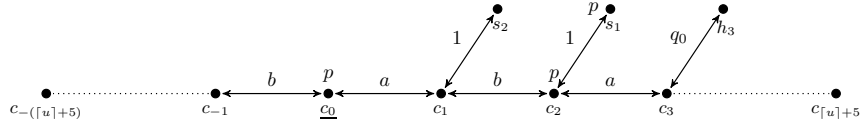
\begin{figure}
\begin{centering}
\scalebox{.75}{
\begin{tikzpicture}[->,>=stealth',shorten >=1pt,shorten <=1pt,auto,node distance=2.8cm,semithick,label distance=0cm]

\tikzstyle{every state}=[draw=black,fill=black,inner sep=0,minimum size=.15cm]

 \node[state,,label=above left:{$$},label=below:{\small $\textcolor{black}{c_{-1}}$}] at (-3,0) (-1) {};
    \node[state,,label=above:{$p$},label=below:{\small $\textcolor{black}{\underline{c_{0}}}$}] at (-1,0) (0) {};
    \node[state,,label=below:{$$},label=below:{\small $\textcolor{black}{c_{1}}$}] at (1,0) (1) {};
    \node[state,,label=above:{$p$},label=below:{\small $\textcolor{black}{c_{2}}$}] at (3,0) (2) {};
    \node[state,,label=below:{$$},label=below:{\small $\textcolor{black}{c_{3}}$}] at (5,0) (3) {};
     \node[state,,label=left:{$p$},label=below:{\small $\textcolor{black}{\ s_{1}}$}] at (4,1.5) (V2) {};
     \node[state,,label=left:{$$},label=below:{\small $\textcolor{black}{\ s_{2}}$}] at (2,1.5) (V1) {};
    \node[state,,label=below:{$$},label=below:{\small $\textcolor{black}{\ h_{3}}$}] at (6,1.5) (S1) {};
   
      \node[state,label=below:{\small $\textcolor{black}{c_{-(\left\lceil u\right\rceil +5)}}$}] at (-6,0) (-2) {};
     \node[state,label=below:{\small $\textcolor{black}{c_{\left\lceil u\right\rceil +5}}$}] at (8,0) (4) {};

\path (-1) edge[<->, draw=black] node [midway] {$b$}
 (0);
\path (0) edge[<->, draw=black] node [midway] {$a$}
 (1);
 \path (1) edge[<->, draw=black]node [midway] {$b$}(2);
 \path (2) edge[<->, draw=black] node [midway] {$a$}(3);
  \path (2) edge[<->, draw=black] node [midway] {$1$}(V2);
   \path (1) edge[<->, draw=black] node [midway] {$1$}(V1);
     \path (3) edge[<->, draw=black] node [midway] {$q_{0}$}(S1);
 \path (3) edge[-,dotted, draw=black](4);
  \path (-1) edge[-,dotted, draw=black](-2);

\end{tikzpicture}
}
\par\end{centering}
\caption{{\small{}\label{fig1}An emulation of a Turing machine in state $q_{0}$
with the read/write head in position 3. Cells 1 and 2 are marked with
$1$ (or $A$), cells -1, 0 and 3 are not.}}

\end{figure}

Let $\mathsf{C}=\{[\mathfrak{c}]\colon\mathfrak{c}\in\mathfrak{C}\}$
be the set of equivalence classes of finite representations of configurations
for which the read-write head is on the finite tape for $\mathsf{M}$
and let $\mathsf{c}=(t,i,q)\in\mathsf{C}$. We construct a pointed
Kripke model $\mathtt{K}(\mathsf{c})$ representing $(t,i,q)$. We
exemplify the construction to be in Fig. \ref{fig1}.

First, in three steps, we construct the set of worlds: $i)$ Construct
slightly too many ``tape cells'': Let $\left\lceil u\right\rceil =\max\{|k|,|k'|\}$
if this is even, else let $\left\lceil u\right\rceil =\max\{|k|,|k'|\}+1$
and take a set of worlds $C=\{c_{j}\colon-(\left\lceil u\right\rceil +5)\leq j\leq\left\lceil u\right\rceil +5\}$.
$ii)$ Represent the content of a cell: Add worlds $S=\{s_{j}\colon\gamma_{j}=1\}$
to indicate ``cells'' with the unique non-blank ``symbol'' 1.
Let $iii)$ Add a ``read/write head'': Let $H=\{h_{j}\colon j=\nicefrac{r}{w}\}$.
Finally, we define the set of worlds as $W=C\cup S\cup H$. 

Second, we add relations between the worlds, also in three steps.
In the following let $R^{*}$ denote the reflexive, symmetric, and
transitive closure of the relation $R$ on a given base set, here
$W$. In particular $(w,w)\in R^{*}$ for all $w\in W$. $i)$ We
structure the cells $c_{i}$ into a tape using relations $R_{a}$
and $R_{b}$: $R_{a}=\{(c_{j},c_{j+1})\colon j\text{ is even}\}^{*}$,
$R_{b}=\{(c_{j},c_{j+1})\colon j\text{ is odd}\}^{*}$. $ii)$ We
attach the non-blank symbols to the appropriate cells: Let $R_{1}=\{(c_{j},s_{j})\colon s_{j}\in S\}^{*}$.
$iii)$ We mount the read/write head at the correct position and in
the correct state, $q$: Let $R_{q}=\{(c_{j},h_{j})\colon h_{j}\in H\}^{*}$.
For the remaining states $q'\in Q\backslash\{q\}$, let $R_{q'}=\{\}^{*}$.
Finally, let $\left\llbracket p\right\rrbracket =\{c_{j},s_{j,},h_{j}\in C\cup S\cup H\colon j\text{ is even}\}$
and the actual world be $c_{0}$. 

We thus obtain a pointed Kripke model $\mathtt{K}(\mathsf{c})=(W,\{R_{i}\}_{i\in Q'},\left\llbracket \cdot\right\rrbracket ,c_{0})$
for the finite configuration representation $\mathsf{c}$ of Turing
machine $\mathsf{M}$. Figure \ref{fig1} illustrates this, depicting
the model $\mathtt{K}(\mathsf{c})$ for configuration $\mathsf{c}=(t,3,q_{0})$.
Given $\mathtt{K}(\mathsf{c})$, we may clearly invert the construction
process and re-obtain an element of $[\mathsf{c]}$.%
{} Finally let $\mathcal{C}=\{\mathtt{K}(\mathsf{c})\colon\mathsf{c}\in\mathsf{C}\}$.

\subsubsection{Expressible Properties.}

To construct an epistemic program that simulates $\delta:\mathsf{C}\rightarrow\mathsf{C}$,
i.e., satisfies Eq. \eqref{this}, we take advantage of the fact that
various properties of configurations are modally expressible. Hence,
we can use these as preconditions. The relevant properties and formulas
are summarized in Table \ref{table}.
\begin{table}
\begin{centering}
\begin{tabular}{p{0.6\textwidth}p{0.01\textwidth}p{0.47\textwidth}}
\textbf{\small{}Property } &  & \textbf{\small{}Formula}\tabularnewline
\hline 
\noalign{\vskip2pt}
{\small{}Being a cell$^{\dagger}$ } &  & {\small{}$c:=(\lozenge_{a}p\vee\lozenge_{b}p)\wedge\lozenge_{a}\neg p$}%
\tabularnewline
\hline 
\noalign{\vskip2pt}
Being a $1$ symbol &  & $s:=\neg c\wedge\lozenge_{1}c$\tabularnewline
\hline 
\noalign{\vskip2pt}
{\small{}Being a cell with symbol $1$} &  & {\small{}$1:=c\wedge\lozenge_{1}\neg c$}\tabularnewline
\hline 
\noalign{\vskip2pt}
{\small{}Being a cell with symbol $0$} &  & {\small{}$0:=c\wedge\neg\lozenge_{1}\neg c$}\tabularnewline
\hline 
\noalign{\vskip2pt}
{\small{}Being the cell of the $\nicefrac{r}{w}$-head is while the
machine is in state $q$} &  & {\small{}$h_{j}:=c\wedge\lozenge_{q}\neg c$}\tabularnewline
\hline 
\noalign{\vskip2pt}
{\small{}Being the cell immediately left of the $\nicefrac{r}{w}$-head
while the machine is in state $q$} &  & {\small{}$l_{q}:=c\wedge\neg h_{q}\wedge\left(\left(p\wedge\lozenge_{a}h_{q}\right)\vee\left(\neg p\wedge\lozenge_{b}h_{q}\right)\right)$}\tabularnewline
\hline 
\noalign{\vskip2pt}
{\small{}Being the cell immediately right of the $\nicefrac{r}{w}$-head
while the machine is in state $q$} &  & {\small{}$r_{j}:=c\wedge\neg h_{q}\wedge\left(\left(p\wedge\lozenge_{b}h_{q}\right)\vee\left(\neg p\wedge\lozenge_{a}h_{q}\right)\right)$}\tabularnewline
\hline 
\noalign{\vskip2pt}
{\small{}Being the cell of/im. left of/im. right of the $\nicefrac{r}{w}$-head
while the machine is in state $j$ and the cell of the $\nicefrac{r}{w}$-head
contains a $1$/$0$} &  & {\small{}$h_{q1}/l_{q1}/r_{q1}/h_{q0}/l_{q0}/r_{q0}:$}\newline
{\small{}Replace $c$ in $h_{q}/l_{q}/r_{q}$ with formula $1/0$.}\tabularnewline
\hline 
\noalign{\vskip2pt}
{\small{}Being a cell at least two cells away from the $\nicefrac{r}{w}$-head} &  & {\small{}$h_{\geq2}:=c\wedge\bigwedge_{q\in Q}\left(\neg h_{q}\wedge\neg l_{q}\wedge\neg r_{q}\right)$}\tabularnewline[2pt]
\hline 
\noalign{\vskip2pt}
{\small{}Being the rightmost$^{\dagger}$ cell } &  & {\small{}$R:=c\wedge\square_{b}\neg p$}\tabularnewline
\hline 
\noalign{\vskip2pt}
{\small{}Being the leftmost$^{\dagger}$ cell } &  & {\small{}$L:=c\wedge\square_{a}\neg p$}\tabularnewline
\hline 
\noalign{\vskip2pt}
{\small{}Being the penultimate cell to the right} &  & {\small{}$PR=c\wedge\neg R\wedge\Diamond_{a}R$}\tabularnewline
\hline 
\noalign{\vskip2pt}
{\small{}Being the penultimate cell to the right} &  & {\small{}$PL=c\wedge\neg L\wedge\Diamond_{b}L$}\tabularnewline
\hline 
\noalign{\vskip2pt}
{\small{}Being at least two steps away from the $\nicefrac{r}{w}$-head
and not being the left- or rightmost cell} &  & {\small{}$2_{AM}:=h_{\geq2}\wedge\neg R\wedge\neg L$ }\tabularnewline
\hline 
\noalign{\vskip2pt}
\end{tabular}
\par\end{centering}
\medskip{}

\caption{\label{table}Expressible properties used as preconditions. \textbf{Notes.}
{\small{}$\dagger$: Recall that the extreme states of $C$ are $c_{-(\left\lceil u\right\rceil +5)}$
and $c_{\left\lceil u\right\rceil +5}$ with $\left\lceil u\right\rceil $
even.}}
\end{table}

\subsubsection{Epistemic Program.}

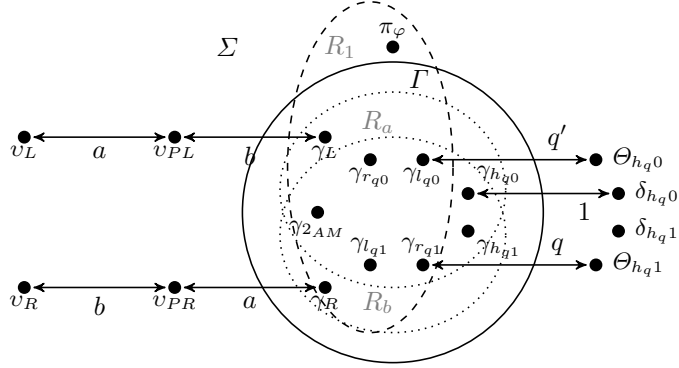
\begin{figure}
\begin{centering}
\begin{tikzpicture}[->,>=stealth',shorten >=1pt,shorten <=1pt,auto,node distance=2.8cm,semithick,label distance=-.1cm]

\tikzstyle{every state}=[draw=black,fill=black,inner sep=0,minimum size=.15cm]

\draw (0,0) circle [x radius=2cm, y radius=2cm];
\draw[dotted] (0,0.3) circle [x radius=1.5cm, y radius=1.3cm];
\draw[dotted] (0,-0.3) circle [x radius=1.5cm, y radius=1.3cm];
\draw[dashed] (-0.3,0.6) circle [x radius=1.1cm, y radius=2.2cm];

    \node at (-2.2,2.2) () {$\Sigma$};
   \node at (0.35,1.77) () {$\Gamma$};
      \node at (-0.2,1.2) () {$\textcolor{gray}{R_{a}}$};
         \node at (-0.2,-1.2) () {$\textcolor{gray}{R_{b}}$};
 \node at (-0.7,2.2) () {$\textcolor{gray}{R_{1}}$};

 \node[state,label=below:{{\small $\textcolor{black}{\gamma_{R}}$}}] at (-0.9,-1) (-1) {};
  \node[state,label=below:{{\small $\textcolor{black}{\gamma_{L}}$}}] at (-0.9,1) (1) {};
 \node[state,label=below:{{\small $\textcolor{black}{\upsilon_{PR}}$}}] at (-2.9,-1) (-2) {};
  \node[state,label=below:{{\small $\textcolor{black}{\upsilon_{PL}}$}}] at (-2.9,1) (2) {};
   \node[state,label=below:{{\small $\textcolor{black}{\upsilon_{R}}$}}] at (-4.9,-1) (-3) {};
  \node[state,label=below:{{\small $\textcolor{black}{\upsilon_{L}}$}}] at (-4.9,1) (3) {};
\node[state,label=above:{{\small $\textcolor{black}{\pi_{\varphi}}$}}] at (0.0,2.2)  {};
  
  \node[state,label=below:{{\small $\textcolor{black}{\gamma_{2_{AM}}}$}}] at (-1,0) (2MI) {};

  \node[state,label=below:{{\small $\textcolor{black}{\gamma_{l_{q0}}}$}}] at (0.4,0.7) (LI) {};
  \node[state,label=below:{{\small $\textcolor{black}{\gamma_{r_{q0}}}$}}] at (-0.3,0.7) (RI) {};

 \node[state,label=above:{{\small $\textcolor{black}{\gamma_{r_{q1}}}$}}] at (0.4,-0.7) (RII) {};
  \node[state,label=above:{{\small $\textcolor{black}{\gamma_{l_{q1}}}$}}] at (-0.3,-0.7) (LII) {}; 

  \node[state,label=above right:{{\small $\textcolor{black}{\gamma_{h_{q0}}}$}}] at (1,.25) (CI) {};
  \node[state,label=below right:{{\small $\textcolor{black}{\gamma_{h_{q1}}}$}}] at (1,-0.25) (CII) {};
  
  
     \node[state,label=right:{{\small $\textcolor{black}{\ \Theta_{h_q 0}}$}}] at (2.7,0.7) (CSI) {};
\node[state,label=right:{{\small $\textcolor{black}{\ \Theta_{h_q 1}}$}}] at (2.7,-0.7) (CSII) {};

 \node[state,label=right:{{\small $\textcolor{black}{\ \delta_{h_q 0}}$}}] at (3,0.25) (DSI) {};
 \node[state,label=right:{{\small $\textcolor{black}{\ \delta_{h_q 1}}$}}] at (3,-0.25) (DSII) {};


\path (-1) edge[<->, draw=black] node [midway] {$a$}(-2);
\path (1) edge[<->, draw=black] node [midway] {$b$}(2);
\path (-2) edge[<->, draw=black] node [midway] {$b$}(-3);
\path (2) edge[<->, draw=black] node [midway] {$a$}(3);

\path (CI) edge[<->, draw=black] node [midway,pos=0.8,below] {$1$}(DSI);

\path (LI) edge[<->, draw=black] node [pos=0.8] {$q'$}(CSI);
\path (RII) edge[<->, draw=black] node [midway,pos=0.8] {$q$}(CSII);

\end{tikzpicture} 
\par\end{centering}
\caption{An illustration of the epistemic program $(\Sigma,\Gamma)$ for a
Turing machine with t $\delta(q,0)=(q',1,l)$ and $\delta(q,1)=(q,0,r)$.
That $\Theta_{h_{q}0}R_{q'}\gamma_{l_{q0}}$ ensures that on input
$(q,0)$ the $\nicefrac{r}{w}$-head moves to the left and the machine
is set to state $q'$ and the relation $\gamma_{h_{q}0}R_{1}\delta_{h_{q}0}$
ensures that the content of the current cell is set to $1$. Similarly
that $\Theta_{h_{q}1}R_{q}\gamma_{r_{q1}}$ ensures that on input
$(q,1)$ the $\nicefrac{r}{w}$-head moves to the right, the machine
remains in state $q$ and the absence of relation $\gamma_{h_{q}1}R_{1}\delta_{h_{q}1}$
ensures that the content of the current cell is set to $0$.}
\label{Figevent} 
\end{figure}

We construct an epistemic program $\Sigma{\scriptstyle \Gamma}=(\Sigma,\{R_{j}\}_{j\in Q'},pre,\Gamma)$
that simulates $\delta:\mathsf{C}\rightarrow\mathsf{C}$, cf. Eq.
\eqref{this}. An example of such an epistemic program is illustrated
in Fig. \ref{Figevent}. 
 We argue for the adequacy of the epistemic program in parallel with
its construction. In the following, the precondition of action $\sigma_{\varphi}$
is the formula $\varphi$.

\subsubsection{Actual actions, halting, and tape enlargement.}

Let the set of actual actions be given by $\Gamma=\{\gamma_{\varphi}\colon\varphi\in\Phi\}$
with $\Phi=\{R,L,2_{AM}\}\cup\{h_{qi},l_{qi},r_{qi}\colon q\in Q,i\in\{0,1\}\},$
cf. Table \ref{table}. 

Then, for any $\mathtt{K}(\mathsf{c})\in\mathcal{C}$, for every cell
state $c_{j}\in C$ of $\mathtt{K}(\mathsf{c})$, $c_{j}$ will satisfy
exactly one of the formulas in $\Phi$. $\Sigma{\scriptstyle \Gamma}$
is thus deterministic over $\mathcal{C}$, and the actual world of
$\mathtt{K}(\mathsf{c})\otimes\Sigma{\scriptstyle \Gamma}$ is a cell.
Finally, formulas from $\Phi$ are only satisfied at cell states of
$\mathtt{K}(\mathsf{c})$. Jointly, this implies that $\Gamma$ ``copies''
the set of tape cells from $\mathtt{K}(\mathsf{c})$ to $\mathtt{K}(\mathsf{c})\otimes\Sigma{\scriptstyle \Gamma}$.

\begin{figure}
\begin{centering}
\scalebox{.75}{
\begin{tikzpicture}[->,>=stealth',shorten >=1pt,shorten <=1pt,auto,node distance=2.8cm,semithick,label distance=0cm]

\tikzstyle{every state}=[draw=black,fill=black,inner sep=0,minimum size=.15cm]

 \node[state,label=above:{$p$},label=below:{{\small $\textcolor{black}{\underline{(c_{0},\gamma_{2_{AM}})}}$}}] at (-3,0) (-1) {};
    \node[state,label=below:{{\small $\textcolor{black}{{(c_{1},\gamma_{2_{AM}})}}$}}] at (-1,0) (0) {};
    \node[state,label=above:{$p$},label=below:{{\small $\textcolor{black}{(c_{2},\gamma_{l_{q_{0}0}})}$}}] at (1,0) (1) {};
    \node[state,,label=below:{{\small $\textcolor{black}{(c_{3},\gamma_{h_{q_{0}0}})}$}}] at (3,0) (2) {};
    \node[state,label=above:{$$},label=below:{\small $\textcolor{black}{(c_{\left\lceil u\right\rceil +5},\gamma_R)}$}] at (5,0) (3) {};
    \node[state,label=above:{$p$},label=below:{{\small $\textcolor{black}{(c_{\left\lceil u\right\rceil +4},\nu_{PR})}$}}] at (7,0) (4) {};
        \node[state,label=below:{{\small $\textcolor{black}{(c_{\left\lceil u\right\rceil +5},\nu_{R})}$}}] at (9,0) (5) {};
      \node[state,label=below:{{\small $\textcolor{black}{(c_{-\left\lceil u\right\rceil -5},\nu_{L})}$}}] at (-6,0) (-2) {};

\path (-1) edge[<->, draw=black] node [midway] {$b$}
 (0);
\path (0) edge[<->, draw=black] node [midway] {$a$}
 (1);
 \path (1) edge[<->, draw=black] node [midway] {$b$}(2);
 \path (3) edge[<->, draw=black] node [midway] {$b$}(4);
 \path (4) edge[<->, draw=black] node [midway] {$a$}(5);
 \path (2) edge[-,dotted, draw=black](3);
  \path (-1) edge[-,dotted, draw=black](-2);

\end{tikzpicture}
}
\par\end{centering}
\caption{\label{fig:tape-longer}Illustration of the extended tape resulting
from applying $\Sigma{\scriptstyle \Gamma}$ to the model in Figure \ref{fig1}.}

\end{figure}
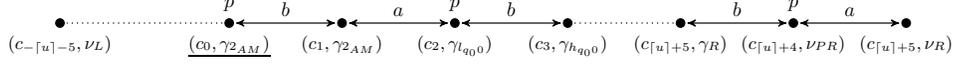

The copied over tape may not be long enough for future operations,
so we include a set of actions to preemptively enlarge it.\footnote{To save tape, this could be done in a more economical manner, only
creating extra cells where actually needed. } To this end, let $\Upsilon=\{\upsilon_{L},\upsilon_{PL},\upsilon_{R},\upsilon_{PR}\}$.
The precondition $\varphi$ of each $\upsilon_{\varphi}\in\Upsilon$
is satisfied by exactly one state $c_{j}$ of $\mathtt{K}(\mathsf{c})$
which is a cell state. These cell state will thus have two successors
in $\mathtt{K}(\mathsf{c})\otimes\Sigma{\scriptstyle \Gamma}$: $(c_{j},\gamma_{\varphi})$
defined before and $(c_{j},\upsilon_{\varphi})$. We thus gain four
new cell states. Setting

\begin{align*}
R_{a} & =\{(\gamma_{\varphi},\gamma_{\psi})\colon\varphi,\psi\in\Phi\backslash\{L\}\}^{*}\cup\{(\upsilon_{PR},\upsilon_{R}),(\gamma_{L},\upsilon_{PL})\}^{*}\\
R_{b} & =\{(\gamma_{\varphi},\gamma_{\psi})\colon\varphi,\psi\in\Phi\backslash\{R\}\}^{*}\cup\{(\gamma_{R},\upsilon_{PR}),(\upsilon_{PL},\upsilon_{L})\}^{*}
\end{align*}
copies over the tape structure and suitably extends it to the new
cell states, which are as the left most, penultimate left, penultimate
right, and right most tape cells. Fig. \ref{fig:tape-longer} illustrates.

\subsubsection{Symbol transfer.}

We copy all symbols from the old tape to the new, safe for the symbol
at the current position of the $\nicefrac{r}{w}$-head. To this end,
add an action $\pi_{\varphi}$ with $\varphi=s\wedge\neg\lozenge_{1}(\bigvee_{\Diamond q\in Q}h_{q})$.
The formula $\varphi$ is then satisfied in $\mathtt{K}(\mathsf{c})$
exactly at the symbols states $s_{j}\in S$ on which the $\nicefrac{r}{w}$-head
is not. Let $\Gamma'=\{\gamma_{\varphi}\colon\varphi\in\Phi\}$ with
$\Phi=\{R,L,2_{AM}\}\cup\{l_{qi},r_{qi}\colon q\in Q,i\in\{0,1\}\}$.
 Requiring that $(\Gamma'\times\{\pi_{\varphi}\})^{*}\subseteq R_{1}$
ensures that the symbol states copied over to $\mathtt{K}(\mathsf{c})\otimes\Sigma{\scriptstyle \Gamma}$
are connected to the correct cell world. We give the precise definition
of $R_{1}$ below.

\subsubsection{Symbol writing.}

We implement the symbol writing part of the transition function $\delta$.
Define a new set of actions by
\[
\Delta=\{\delta_{h_{q}i}\colon q\in Q,i\in\{0,1\}\text{ and }\delta(i,q)\text{ is defined}\}.
\]
At most one action from $\Delta$ will have its precondition satisfied
at any $\mathtt{K}(\mathsf{c})$ and just in case $\delta(\mathsf{c})$
is defined. The world satisfying this precondition is a cell world,
$c_{j}$, which will have two successors in $\mathtt{K}(\mathsf{c})\otimes P$:\footnote{Possibly three, see below.}
a cell world successor $(c_{j},\gamma_{h_{q}i})$ defined above and
a symbol world successor $(c_{j},\delta_{h_{q}i})$ defined here.
We ensure that the emulation writes the correct symbol by connecting
$(c_{j},\delta_{h_{q}i})$ to $(c_{j},\gamma_{h_{q}i})$ by $R_{1}$
or not: Let
\[
R_{tmp}=\{\{(\delta_{h_{q}i},\gamma_{h_{q}i})\colon\gamma\in\Gamma\}\ |\ \delta(i,q)=(\cdot,1,\cdot)\}
\]
and let $R_{1}=((\Gamma^{'}\times\{\pi_{\varphi}\})\cup R_{tmp})^{*}$.
This and the above ensures that the emulation produces a correctly
labeled tape. 

\subsubsection{State change and head repositioning.}

We finally implement the state change and head repositioning encoded
by $\delta$. To this end, define a set of events
\[
\Theta=\{\theta_{h_{q}i}\colon q\in Q,i\in\{0,1\}\text{ and }\delta(i,q)\text{ is defined}\}.
\]
Again, at most one action from $\Theta$ will have its precondition
satisfied at any $\mathtt{K}(\mathsf{c})$ and just in case $\delta(\mathsf{c})$
is defined. The world satisfying this precondition is a cell world,
$c_{j}$, which will hence have two successors in $\mathtt{K}(\mathsf{c})\otimes\Sigma{\scriptstyle \Gamma}$:\emph{}\footnote{Possibly three, cf. the above.}
a cell world successor $(c_{j},\gamma_{\varphi})$ defined above and
a $\nicefrac{r}{w}$-head world successor $(c_{j},\theta_{h_{q}i})$
defined here. We ``mount'' the $\nicefrac{r}{w}$-head world at
the correct position and in the correct state using the relations
$\{R_{q'}\}_{q'\in Q}$: For all $q'\in Q$, let 
\[
R_{q'}=\{(\gamma_{xq'},\theta_{h_{q}i})\colon\delta(q,i)=(q',\cdot,x),i\in\{0,1\},q\in Q\}^{*}.
\]

The definition of $\{R_{q}\}_{q\in Q}$ ensures that the $\nicefrac{r}{w}$-head
is moved and changes state appropriately, whenever $\delta(i,q)$
is defined. When $\delta(i,q)$ is not defined, the $\nicefrac{r}{w}$-head
world $(c_{j},\theta_{h_{q}i})$ will be disconnected from the tape
cell worlds. In that case, $\mathtt{K}(\mathsf{c})\otimes P$ will
not be in $\mathcal{C}$, and the emulation is said to halt. This
concludes the construction and proof.\vspace{-1.5\baselineskip}
\begin{flushright}\textsf{QED}\end{flushright}
\begin{remark}
The proof generalizes to $k$-tape Turing machines or bigger input
symbol sets by replacing modality $\Box_{1}$ with $\Box_{1},\Box_{k}$
and the corresponding formula $1$ with $1,..,k$. 
\end{remark}

\bibliographystyle{abbrv}

\end{document}